\documentclass [12pt]{article}
\usepackage {amssymb}
\usepackage {amsmath}
\usepackage [cp1251]{inputenc}
\usepackage{graphicx}
\usepackage {longtable}

\title{Interaction of a dipole point vortex with flat boundary}

\date{}

\author{ \textbf{A.V.Tur},
\textbf{V.V.Yanovsky}}
\begin{document}

 \maketitle
$^{\ast}$\textit{Institut de Recherche en Astrophysique et Plan\'{e}tologie C.N.R.S.-U.P.S., 9, avenue Colonel-Roche 31028 Toulouse Cedex 4, France.}

\textit{Institute for Single Crystals, Ukranian National Academy of Science, Kharkov 61001, Ukraine}

\abstract{In this work we have found an exact solution for the problem of the movement of a dipole type point vortex in an area of fluid limited by a flat boundary. We also present a solution to  the problem of dipole point vortex motion in a right angle. It is shown that unlike a usual point vortex, the dipole vortex always comes away from the boundary asymptotically. This important feature of the dipole vortex allows it to be considered to be one of the efficient mechanisms of vorticity transfer from boundary to media}

\section{Introduction}

The new type of dipole point vortex was found in work \cite {dip0}. Also in this work, the motion equation system was obtained and studied for one-dimensional singularities which are compatible with the Euler equation. We point out that the point dipole vortices are weak solutions of Euler equation. It should be noted that point dipole vortices are Euler equation solutions only in cases where dipole moments themselves are evolving over time.This kind of vortex can be considered as a specific hydrodynamic quasiparticle as well as an already-recognized point vortex. The interaction between dipole vortices and usual point vortices defines a field of velocities in two-dimensional hydrodynamics. (Singularities of higher orders are not dynamically compatible with the Euler equation, but they can exist as stationary solutions.) The system of any number of such vortices is hamiltonian and has three integrals of motion in involution. According to the Liouville theorem, this means that the problem of interaction of a usual vortex and the other, of the dipole type \cite {dip0} is exactly integrated in a similar manner, like the known case of integration of system of three usual point vortices \cite {dip0a}, \cite {dip0b}, \cite {dip0c}. Exact solutions for, and the behaviour of two-point vortices, one of which is the point dipole, were considered in work \cite {dip01}. In work \cite {dip02} new stationary solutions with complex singuliarities of two-dimensional ideal hydrodynamics were obtained with help of point dipoles. But in these works, point dipole vortices were considered in fluid without boundaries. In this work, the simplest case of the motion of one point dipole vortex with a flat boundary to media is discussed\footnote{When our work was sent to the Journal we learned that two weeks earlier, work [9] had appeared in which motion equations of the dipole vortex near the plane boundary and solutions for  dipole vortex movement were obtained, independently and using a different method (complex variables) In our work we have used the images method and analyzed in more detail the behavior of a dipole vortex near a plane boundary. Additionally, we studied the behavior of a vortex near a boundary which has the form of a right angle. This strengthens the main conclusion of our work concerning the role of a dipole vortex in efficient vorticity exchange between a boundary and media. }. This problem is exactly integrable using the method of images and gives a complete description of all the modes of motion of a vortex near a flat boundary. In addition, we examine point dipole vortex motion at a right angle. As a result, we have established the simple asymptotic laws of vortex evolution on a long time scale and have found that the dipole vortex always moves away from the boundary over time. This means that we obtained an effective carrier of vorticity from boundary to media. This property is important, since, as a rule, vorticity origination is connected to boundaries.

\section{Search for the image}

\begin{figure}
  \centering
  \includegraphics[width=5 cm]{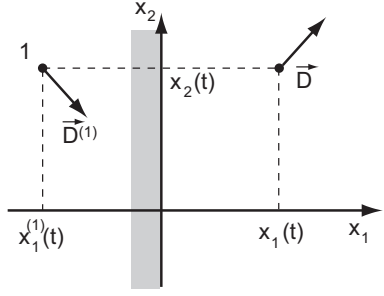}\\
  \caption{Position of the boundary and the point dipole vortex is shown. The fluid occupies semispace $x_1> 0$.}\label{fg1}
\end{figure}

To determine the movement of a dipole point vortex near a solid boundary it is possible to use the method of images \cite {dip1},\cite {dip2}. According to this method, the point dipole vortex at a wall corresponds to the system of two point vortices of a dipole type in fluid without a boundary. Such a system of point vortices must satisfy the same boundary conditions as the initial problem. Let the dipole vortex be at flat boundary   with fluid  $x_1=0$ which fills the semispace  $x_1> 0$ (see Fig.\ref {fg1}). With this geometry, the stream function must satisfy the boundary condition  $ \varphi | _ {x_1=0} =0$. Hence it is necessary to find the conditions when the dipole vortex, interacting with its image, would satisfy the boundary condition described above. The vortex-image settles at points $x_1^{(1)} =-x_1 $  and $x_2^{(1)} =x_2 $  for reasons of symmetry. It remains to find out how the dipole moment of the image $ \vec {D} ^{(1)}$  is connected with the dipole moment of the initial vortex  $ \vec {D} = (D_1, D_2) $. We can note the necessary conditions using the stream function of these two vortices:
\begin{equation}\label{d1}
    \varphi=-\frac{1}{2 \pi} \left. \left\{ D_i (t) \frac{x_i - x_i (t)}{|\vec{x}-\vec{x}(t)  |^2}+ D_i^{(1)} (t) \frac{x_i - x_i^{(1)} (t)}{|\vec{x}-\vec{x}^{(1)}(t)  |^2}\right\} \right|_{x_1=0}=0
\end{equation}
Here $ \vec {x} (t) $  is the position of the initial vortex, and $ \vec {x} (t) ^ {(1)} $  is the position of its image. Taking into account the relation between the vortices' coordinates, this condition transforms easily into
\[-D_1 (t) x_1 (t)+D_2 (t) (x_2-x_2(t)) + D_1^{(1)} (t) x_1 (t)+D_2^{(1)} (t) (x_2-x_2(t))=0\]
and it is satisfied when
\begin{equation}\label{d2}
    D_1^{(1)} (t) =D_1 (t), \qquad D_2^{(1)} (t) = -D_2 (t)
\end{equation}
Then, the behaviour of a vortex near the solid boundary can be described in the boundless media with two point dipole vortices with the special choice of their characteristics.

The component of the image dipole moment which is normal to the boundary coincides with the normal component of the initial vortex and the tangential image component reverses the sign. Using this fact one can consider the behaviour of the dipole vortex for other boundary conditions as well. In particular, it is easy to study the dipole vortex motion in a fluid limited by solid boundary forming a right angle ( the area occupied by fluid $x_1 > 0$ and $x_2 > 0$).

\begin{figure}
  \centering
  \includegraphics[width=6 cm]{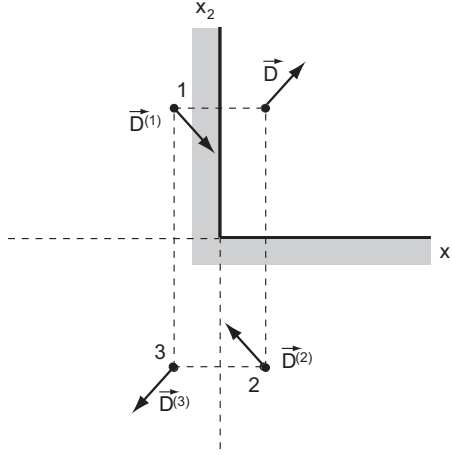}\\
  \caption{Shows the position of the dipole vortex in the point $\vec{x}=(x_1 , x_2)$ with the dipole moment $\vec{D}=(D_1 , D_2)$ in a right angle. The same boundary conditions appear when three other dipole vortices images  are located in an unbounded media with the coordinates $\vec{x}^{(1)}=(-x_1 , x_2)$, $\vec{x}^{(2)}=(x_1 , -x_2)$, $\vec{x}^{(3)}=(-x_1 , -x_2)$ and dipole moments chosen in accordance with $\vec{D}^{(1)}=(D_1 , -D_2)$, $\vec{D}^{(2)}=(-D_1 , D_2)$, $\vec{D}^{(3)}=(-D_1 , -D_2)$. (The superscript in parentheses  corresponds  to the vortex number).}\label{fg2}
\end{figure}

The current function vanishing at the media boundary can be obtained by placing image vortices as shown on Fig.\ref{fg2}. In the case of the right angle boundary three image vortices are sufficient. It is easy to verify that the current function of this dipole vortices configuration vanishes with $x_1 = 0$ and $x_2 = 0$. Consequently, the dipole vortex evolution in a right angle  comes to the problem  of motion of four dipole vortices in boundless media. The values of these dipole vortices moments are shown on Fig.\ref{fg2}. In much the same way, it is easy to obtain vortices motion equations for angles of a different  value as well as for a circular area and other elementary boundaries.

\section{Equations of motion}

To obtain the hamiltonian of this system of vortices we will use the general hamiltonian which was found in work \cite {dip0}. In accordance with this work, the hamiltonian of two interacting dipole vortices is
\begin{equation}\label{d3}
    H=-\frac{1}{2\pi} \frac{2D_{m}
(x_{m} -x_{m}^{(1)})D_{l}^{(1)}
(x_{l}-x_{l}^{(1)})-D_{m}D_{m}^{(1)}(\vec{x}-\vec{x}^{(1)})^2}
 {|\vec{x}-\vec{x}^{(1)}|^4}
\end{equation}
Here $\vec{x}$, $\vec{x}^{(1)}$ -  are the coordinates and $\vec{D}$, $\vec{D}^{(1)}$ - are the dipole moments of two dipole vortices. Now let us consider the relation which was demonstrated earlier between the positions and dipole moments of the vortex and its image. After simple transformations we obtain
\begin{equation}\label{d4}
    H=-\frac{1}{2\pi} \frac{D_1^2 +D_2^2}{4 x_1^2}
\end{equation}
Based on this hamiltonian it is easy to get the equations of motion of the dipole vortex near a solid wall in the form:
\begin{equation}\label{dp4}
    \frac{d x_i}{dt}=-\varepsilon_{i k} \frac{\partial H}{\partial D_k}
\end{equation}
\begin{equation}\label{dp5}
    \frac{d D_i}{dt}=-\varepsilon_{i k} \frac{\partial H}{\partial x_k}
\end{equation}
Where $ \varepsilon _ {i k} $  is the antisymmetric unit tensor. After a substitution of the Hamiltonian (\ref{d4}) we obtain the following system of equations:
\begin{equation}\label{d5}
    \frac{d x_1}{d t}=\frac{D_2}{4 \pi x_1^2}
\end{equation}
\begin{equation}\label{d6}
    \frac{d x_2}{d t}=-\frac{D_1}{4 \pi x_1^2}
\end{equation}
\begin{equation}\label{d7}
    \frac{d D_1}{dt}=0
\end{equation}
\begin{equation}\label{d8}
    \frac{d D_2}{dt}= \frac{D_1^2 +D_2^2}{4 \pi x_1^3}
\end{equation}
Firstly, from this system of equations follows the conservation of the orthogonal to the boundary component of the dipole moment of the vortex  $D_1 (t) = D_1 (0) \equiv const $. This is the consequence of the law of general conservation $I=\sum_{\alpha}\Gamma_{\alpha}x_{v1}^{(\alpha)}- \sum_{\beta} D_1^{(\beta)}$, which is satisfied for the  system  of interacting usual point vortices and  point dipole vortices \cite {dip0}. In this equation $\Gamma_{\alpha}$ is the vortex strength of $\alpha$-point vortex  and $x_{v1}^{(\alpha)}$-its coordinates. Therefore the system splits into two subsystems of equations. It is enough to solve the closed system of two equations (\ref{d5}), (\ref{d8}) to get all the vortex characteristics. It is important to note that the system of equations (\ref{d5})-(\ref{d8}) can be directly obtained from the equations of motion of two dipole vortices
\cite {dip0}, after substitution of the relation between coordinates and the dipole moments\footnote{For the equation system (\ref{d5})-(\ref{d8}) given above we used different variables than those in work \cite{PSE}.}. It is evident that the energy of a point vortex is conserved and therefore the value of the dipole moment is defined by the distance to the boundary
\[-\frac{1}{2\pi} \frac{D_1^2 +D_2^2}{4 x_1^2} =E_0\]
where $E_0$  is the initial value of vortex energy. Then it obviously follows that
\[D_1^2 +D_2^2 =- 8\pi  E_0 x_1^2 \]
We can define the dependence on time of the longitudinal movement to the boundary component of the dipole moment $D_2$  using this invariant. In order to do it we solve this equation with respect to $x_1$
\begin{equation}\label{d8a}
    x_1= + \left(\frac{D_1^2 +D_2^2}{-8\pi  E_0} \right)^{\frac{1}{2}}
\end{equation}
Here we take into account that energy of the dipole vortex $E_0 <0$ and fluid occupies the positive semiplane  $x_1 \geq 0$. This defines the choice of the positive sign of the square root. We exclude coordinate $x_1$   from the equation (\ref {d8}) with help of this formula. As a result we get
\[\frac{d D_2}{dt}= \frac{C}{(D_1^2 +D_2^2)^{\frac{1}{2}}}\]
where the constant  $C = (-8 \pi E_0) ^ {\frac {3} {2}}/4 \pi> 0$ is defined by vortex energy. Taking into account the conservation of $D_1$, we can integrate this equation in elementary functions
\begin{equation}\label{d9}
     D_2 \sqrt{D_1^2 +D_2^2} + D_1^2 \ln \left(D_2 + \sqrt{D_1^2 + D_2^2} \right)= 2 C t +const
\end{equation}
The integration constant is defined by the initial conditions in accordance with the equation  $ const =D_2 (0) \sqrt {D_1^2 +D_2^2 (0)} + D_1^2 \ln \left(D_2 (0) + \sqrt {D_1^2 + D_2^2 (0)} \right)$. Here, and further in this article, for initial values of $D_2 (t)$ we use the designation of the $D_2 (0)$ type. From this solution it is easy to find the asymptotic increase of $D_2 (t)$  according to the relation
\[ D_2 (t) \sim \sqrt{2 C t} \]
Naturally, at the given law of evolution $D_2 (t)$, the change of coordinates of the vortex can be  easily calculated. Then $x_1(t)$  is defined from $D_2 (t)$  by an algebraic equation (\ref {d8a}) which can be written as
\begin{equation}\label{d9a}
    x_1= x_1(0) \left(\frac{D_1^2 +D_2^2 (t)}{D_1^2 +D_2^2 (0)} \right)^{\frac{1}{2}}
\end{equation}
using the law of energy conservation. After that, the coordinate   is obtained by  the integration of the equation (\ref {d6}). The degeneration solution with $D_1 =0$ was considered in article \cite{PSE}. In this case in accordance with equation (\ref{d9a}) the position of the dipole vortex $x_1 =D_2 \frac{x_1(0)}{D_2 (0)}$ is proportional to the non-zero dipole moment component, and the solution of equation (\ref{d8}) takes the evident form $D_2^2 (t)=D_2 (0)^2 \cdot (1+t\frac{D_2(0)}{2 \pi x_1 (0)^3}) $. It follows that with $D_2(0) < 0$ the dipole momentum vanishes over a finite time and consequently only in this case the dipole vortex approaches the boundary. With any small deviation from zero of the normal to the boundary dipole moment  $D_1$ component, the approach stage to the boundary switches after  several times to move away from boundary to infinity.

\section{Movement of vortex at solid boundary}

\begin{figure}
  \centering
  \includegraphics[width=7 cm]{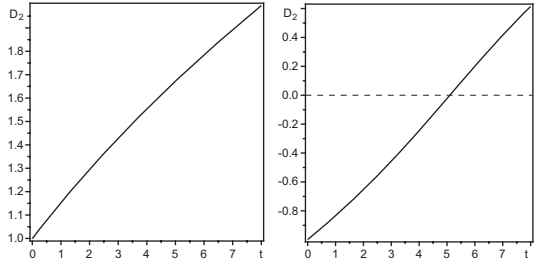}\\
  \caption{Change of $D_2$  in time: on the left for initial conditions $D_2 (0) =1$, $D_1 (0) =1$, $x_1 (0) =1$  and  $x_2(0) =0$, on the right  $D_2(0) =-1$ and other initial data coincide with the previous initial conditions.}\label{fg3}
\end{figure}

Let us now consider the character of the movement of the vortex at the wall with help of the previously-obtained exact solutions. First of all, one can notice, that the component of the dipole moment along the wall increases with time. Over long times it follows the square root law. At the initial stage at $D_2(0)<0$  this component decreases at first, and then increases (see Fig.\ref {fg3}).

\begin{figure}
  \centering
\includegraphics[width=8 cm]{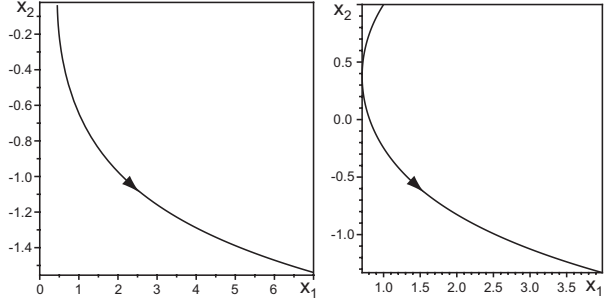}\\
  \caption{On the left is shown the trajectory of vortex movement for  $D_1> 0$ and  $D_2> 0$ and on the right  for  $D_1> 0$ and  $D_2 <0$. The direction of movement is marked by the arrow on the trajectory. On the right, one can see the beginning stage of  the approach to the boundary and further moving away from it.}\label{fg4}
\end{figure}

As a result of this behaviour of the longitudinal component of the dipole moment, over long times the vortex moves asympoticaly away from boundary as
\begin{equation}\label{d10}
    x_1 \approx x_1 (0) \left(\frac{2 C t}{D_1^2 + D_2^2(0)} \right)^{\frac{1}{2}}
\end{equation}
Hence, for $D_2 (0) <0$  at the beginning, the vortex is approaching the boundary at a minimal distance (see eq. (\ref {d9a}) )
\[x_{min}= x_1(0) \frac{|D_1 |}{(D_1^2 +D_1^2 (0))^{\frac{1}{2}}}\]
Then the vortex moves away from the boundary asymptotically in accordance with formula (\ref {d10}). Now we have to discuss the movement of vortex along the boundary. For this purpose we will come back to equation (\ref {d6}). The equations (\ref {d5}) and (\ref {d8a}) give:
\[\frac{dx_1}{dx_2}=-\frac{\sqrt{(D_1^2 +D_2^2 (0))x_1^2 -x_1(0)^2 D_1^2}}{D_1 x_1(0)}\]
Integrating this equation we obtain the vortex movement trajectory along the boundary:
\begin{equation}\label{d11}
    \frac{\ln \left({\sqrt{D_1^2 +D_2^2 (0)} x_1} + \sqrt{(D_1^2 +D_2^2 (0))x_1^2 - D_1^2 x_1(0)^2}  \right)}{{\sqrt{D_1^2 +D_2^2 (0)}}}=-sign(D_2 (0))\frac{x_2}{D_1 x_1(0)} + const
\end{equation}
The integration constant is defined by initial conditions. Examples of the trajectories of vortices' movement are shown in Fig. \ref {fg4}.

The main conclusion that follows from the equation (\ref{d11}) is that the vortex moves away exponentially from the wall. It is important to note that a usual point vortex moves with constant speed along a flat boundary maintaning the same distance from it. The direction of the longitudinal motion of dipole vortex is defined by the sign of  $D_1 (0)$. From a physical point of view, this means that vortices of a dipole type can generate an effective mechanism of vorticity transfer from the boundary where it is generated to the media. Naturally, the dipole vortices can also intensify the transfer of other passive "impurities", such as temperature to media. For many physical phenomena this is a highly important property.

In conclusion, we must highlight, that the case of two interacting dipole vortices is integrated in quadratures with the special choice of dipole moments. This choice is connected with zero values of some first integrals of motion. The case considered above corresponds to zero value of $I=\sum D^{\alpha}_2 =0$  and  $J=\sum \vec{D}^{\alpha} \cdot \vec{x}^{\alpha} =0$. The remaining integral $\sum D^{\alpha}_1=const$  and energy can take any value. In general cases, the problem of dynamics of two point dipole vortices is not integrated.

\section{Movement of dipole vortex in a right angle}

Using the images method one can obtain the hamiltonian of the dipole vortex in media limited by	 a right angle. With help of the hamiltonian of four interacting dipole vortices and after the substitution of coordinates and dipole moments of images we obtain:
\[H=-\frac{1}{4 \pi} \left\{ \frac{D_1^2 +D_2^2}{x_1^2} +  \frac{D_1^2 +D_2^2}{x_2^2}- \frac{(x_1^2 -x_2^2)(D_1^2 -D_2^2)+4D_1 D_2 x_1 x_2}{(x_1^2 +x_2^2)^2}\right\}\]
Here $x_1 > 0$ and $x_2 > 0$. As earlier, this hamiltonian can be considered to be like the hamiltonian of dipole vortex moving in an angle. We obtain the motion equation using the standard way in accordance with the equations:
\[\frac{d x_1}{d t}=\frac{1}{2 \pi} \left(\frac{D_2}{x_1^2}+\frac{D_2}{x_2^2}+\frac{D_2 (x_1^2 - x_2^2)-2 D_1 x_1 x_2}{(x_1^2 + x_2^2)^2}  \right)\]

\[\frac{d x_2}{d t}=-\frac{1}{2 \pi} \left(\frac{D_1}{x_1^2}+\frac{D_1}{x_2^2}-\frac{D_1 (x_1^2 - x_2^2)+2 D_2 x_1 x_2}{(x_1^2 + x_2^2)^2}  \right)\]

\[\frac{d D_1}{d t}=-\frac{1}{2 \pi} \left(\frac{D_1^2 + D_2^2}{x_2^3}+\frac{(D_2^2 -D_1^2)x_2 +2 D_1 D_2 x_1}{(x_1^2 + x_2^2)^2}-2x_2 \frac{(D_1^2 -D_2^2) (x_1^2 - x_2^2)+ 4 D_1 D_2 x_1 x_2}{(x_1^2 + x_2^2)^3}  \right)\]

\[\frac{d D_2}{d t}=\frac{1}{2 \pi} \left(\frac{D_1^2 + D_2^2}{x_1^3}+\frac{(D_1^2 -D_2^2)x_1 +2 D_1 D_2 x_2}{(x_1^2 + x_2^2)^2}-2x_1 \frac{(D_1^2 -D_2^2) (x_1^2 - x_2^2)+ 4 D_1 D_2 x_1 x_2}{(x_1^2 + x_2^2)^3}  \right)\]
Obviously this is a more complex dynamical  system  with two degrees of freedom. The energy of this dipole vortices system is conserved. However, other conservation laws obtained in work \cite{dip0} become trivial. For  the considered configuration of the dipole vortices they vanish. That is why the question regarding the integrability in quadrature of this equation system remains open.  But it is easy to see  the discrete symmetry of equations which is due to the permutation of coordinates.
\begin{figure}
  \centering
  \includegraphics[width=5 cm]{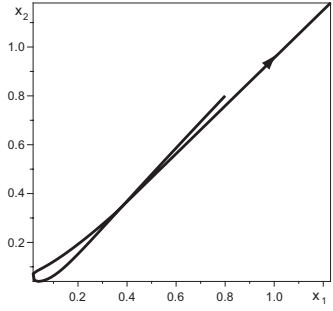}\\
  \caption{Movement of the dipole vortex with the initial coordinates $x_1 (0)=0.8$, $x_2 (0)=0.8$ and initial value of the dipole moment $D_1 (0)=1.1$ $D_2 (0)=-1$, which are obtained numericaly. A similar qualitative modification of vortex movement  appears under any minor breaking of symmetry, for instance with conservation of $D_1 =- D_2$, but with deviation from the diagonal $x_1 (0) \neq x_2(0)$. }\label{fg5}
\end{figure}
This permits us to find the simple particular case  of dipole vortex movement with $x_1 = x_2$ and $D_1 =- D_2$. The motion equations for it are simplified and take the form:
\[\frac{d x_1}{d t}=-\frac{5}{4 \pi} \frac{D_1}{x_1^2}\]

\[\frac{d D_1}{d t}=-\frac{5}{4 \pi} \frac{D_1^2}{x_1^3}\]
This particular case is exactly integrable. And the solution of the equation system has the form:
\[D_1 = \left( \frac{D_1 (0)}{x_1 (0)} \right)x_1\]
\[x_1^2 =x_1^2 (0) -\frac{5}{2}t \frac{D_1 (0)}{x_1 (0)} \]
Hence, with $D_1 (0) > 0$ the dipole vortex approaches the angle vertex, and with initial condition $D_1 (0) < 0$, moves away along the angle bisector from its vertex towards media. The character of dipole vortex movement is sensitive to deviations of the  condition $D_1 =- D_2$. So, even with small deviations of this condition, the dipole vortex approaching the angle vertex begins to move away from it over time (see Fig.\ref{fg5})
\begin{figure}
  \centering
  \includegraphics[width=5 cm]{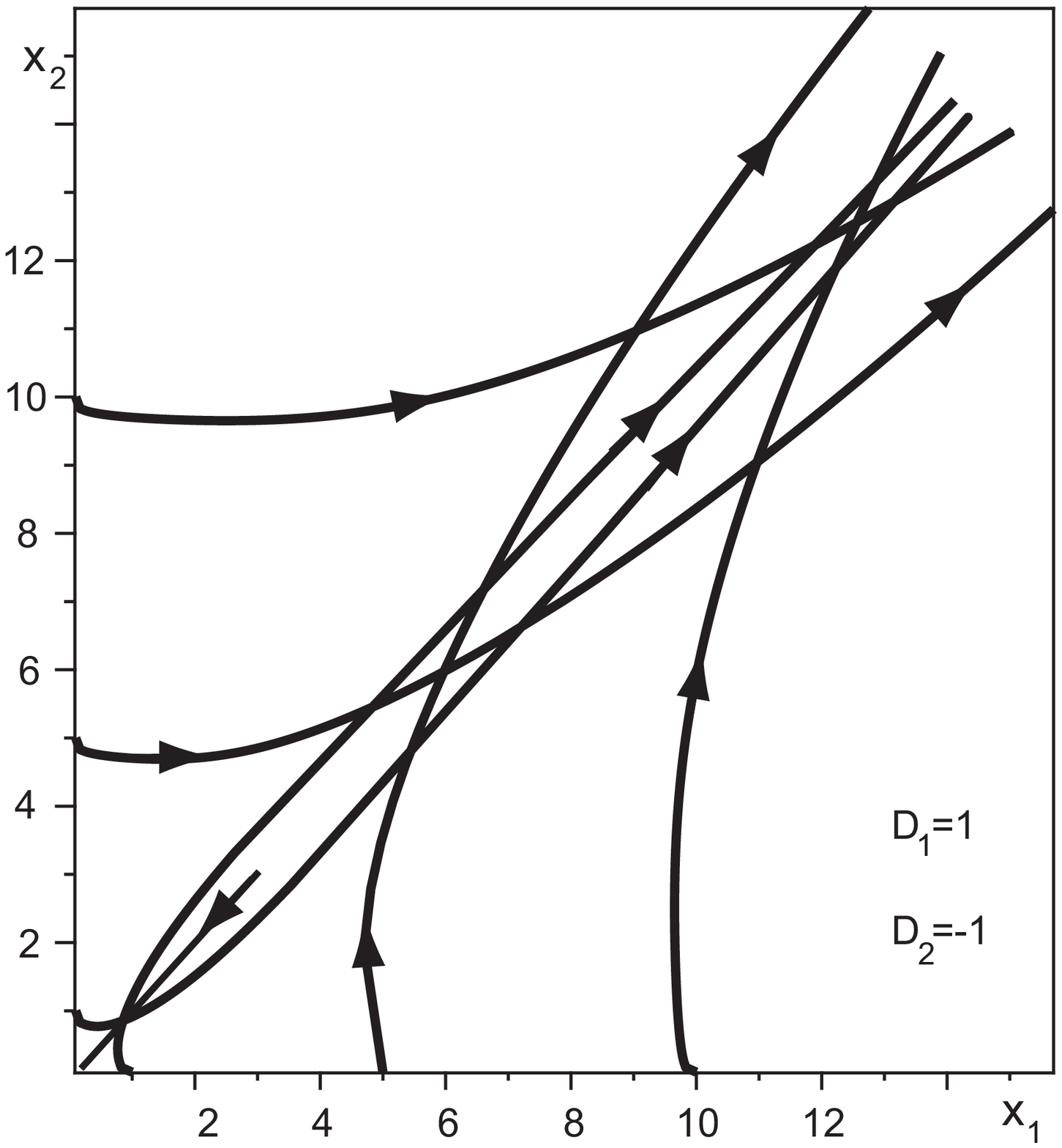}
  \includegraphics[width=5 cm]{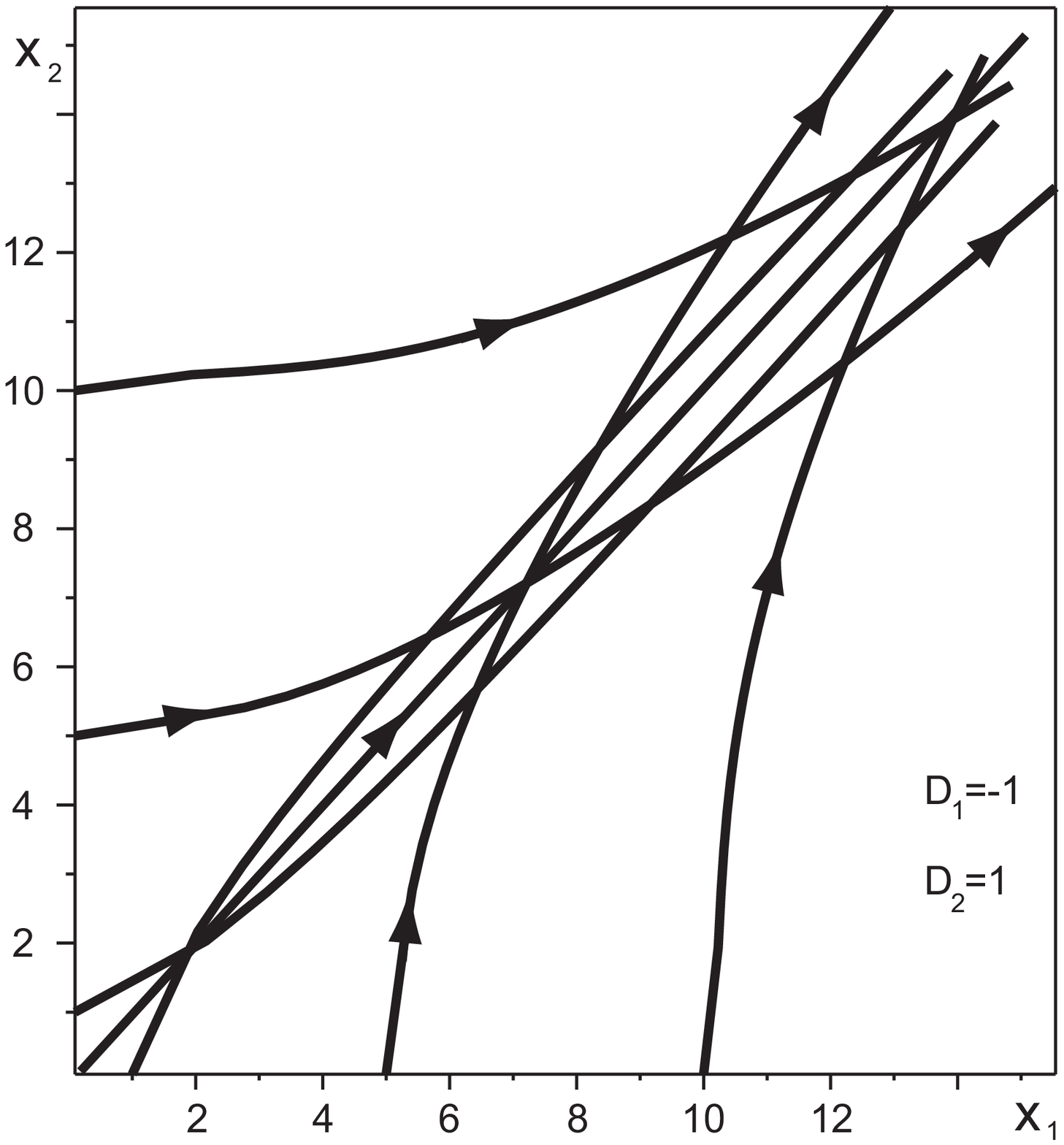}\\
  \caption{Shows the trajectory projections of the movement of vortices on plane $(x_1 , x_2)$with different initial dipole moments, as well as different initial positions of the dipole vortices in the media. On the left $D_1 (0) = 1$, $D_2 (0) =-1$, and on the right $D_1 (0) = -1$, $D_2 (0) =1$.On the left only one trajectory in accordance with the analytical solution obtained above, moves to the angle vertex. On the right, the same trajectory moves away from the angle vertex along the bisector.}\label{fg6}
\end{figure}
We can also examine the asymptotical behaviour of the vortex for example with $x_2 \gg x_1$. In zero approximation, the vortices motion equations system coincides with equations (\ref{d5})-(\ref{d8}). This allows us to describe the asymptotical behaviour of the vortex.  The numerical results also demonstrate that the moving away of dipole vortices from angle vertex to media is typical (see. Fig.\ref{fg6}). Consequently, for the case of a right angle, the dipole performs an efficient exchange of vorticity between boundary and media.

\section{Conclusion}

With the dipole point vortex the set of specific quasi-particle increases. This allows us to consider problems of bi-dimensional hydrodynamics within the framework of finite-dimensional dynamics systems. Hence the number of problems which are needed to take into account the impact of point dipole vortices increases as well. They rank with usual point vortices, but they also have some interesting properties.

On one hand, dipole vortices are an example of unusual hamiltonian systems which appear without using Lagrange’s formalism and Legendre’s transformations. The hamiltonian nature of the dynamical equations of interacting dipole vortices plays an important part in the search of equation solutions as well as for many other problems. New examples of exactly integrable problems appear due to the existence of additional conservation laws. The examples of these exactly integrable systems are given above, but some others are known, for which references were given in the Introduction.
On the other hand, these point dipole vortices have different properties compared to the usual point vortices. In particular, as was shown earlier with the solution of the problem of point dipole movement for an area limited by a plane boundary as well as for an area limited by a right angle,  the important function of the point dipole vortices may be to provide a vorticity exchange mechanism between boundaries and media. This property is very important for many phenomena concerned with vortex movements, because as a rule, vorticity is generated near the boundaries, as in the classical examples of the flow around a body, appearance of von Karman vortex street, wing lift mechanism etc. From this point of view, it is interesting to consider the advection which appears in the field of interacting usual and point dipole vortices. It is evident also that there are more problems concerned with the statistical description of point vortices system. The statistical description of the singularities set stands in close relation to bi-dimensional turbulence. The appearance of dipole point vortices in superfluid liquid and in other complex hydrodynamical mediums is also of great interest, as well as the existence of point dipole vortices on the bi-dimensional manifolds. Most of these problems remain unanswered.

\section*{Acknowledgments}

We would like to thank the anonymous reviewer for drawing our attention to the article  [9], where some results of our work were obtained independently.

\end{document}